\documentclass[aps,prl,twocolumn,showpacs,showkeys,groupedaddress]{revtex4}
\usepackage{graphicx}
\usepackage{epstopdf}
\usepackage{color}
\usepackage{url}
\usepackage{hyperref}
\usepackage[T1]{fontenc}
\usepackage[latin1]{inputenc}
\usepackage{float}
\usepackage{ulem}
\usepackage{amsmath}

\newcommand{\ket}[1]{|{#1}\rangle}

\newcommand{\proj}[2]{\left\vert #1 \right\rangle \!\left\langle #2 \right\vert } 

\newcommand{\Ca}{$^{40}\mathrm{Ca}^+$~}
\newcommand{\e}{\mathrm{e}}
\def\figref{Fig.~}
\let \pt=\partial

\let \mr=\mathrm
\let \mc=\mathcal
\let \eps=\varepsilon
\let \dag=\dagger

\makeatletter
\newcommand*{\rom}[1]{\expandafter\@slowromancap\romannumeral #1@}
\makeatother

\begin{document}
	\title{Strong coupling of a single ion to an optical cavity}
	
	\author{Hiroki Takahashi}\email{takahashi@qc.rcast.u-tokyo.ac.jp}\thanks{Present address: Research Center for Advanced Science and Technology, the University of Tokyo, Meguro-ku, Tokyo 153-8904, Japan}
	\author{Ezra Kassa}
	\author{Costas Christoforou}
	\author{Matthias Keller}
	\affiliation{Department of Physics and Astronomy, University of Sussex, Brighton, BN1 9QH, United Kingdom}
	
	\begin{abstract}
Strong coupling between an atom and an electromagnetic resonator is an important condition in cavity quantum electrodynamics (QED). While strong coupling in various physical systems has been achieved so far, it remained elusive for single atomic ions. In this paper we demonstrate for the first time the coupling of a single ion to an optical cavity with a coupling strength exceeding both atomic and cavity decay rates. We use cavity assisted Raman spectroscopy to precisely characterize the ion-cavity coupling strength and observe a spectrum featuring the normal mode splitting in the cavity transmission due to the ion-cavity interaction. Our work paves the way towards new applications of cavity QED utilizing single trapped ions in the strong coupling regime for quantum optics and quantum technologies.  
		
	\end{abstract}
	\maketitle
	
	Coupling between atoms and electromagnetic fields is a ubiquitous physical process that underlies a plenitude of electromagnetic phenomena. 
	In cavity quantum electrodynamics (QED), this interaction is studied in its simplest form where a single atomic emitter is coupled to the well-defined electromagnetic modes of a resonator. Over the years, cavity QED has grown into a versatile platform for the investigation of fundamental atom-photon interactions and an indispensable tool for quantum information technologies \cite{kimble1998strong,girvin2009circuit}.
	In many applications, the coherent atom-photon interaction rate needs to exceed the decoherence rates of the system. This so-called strong coupling regime has been attained in many physical systems including neutral atoms \cite{boca2004observation,maunz2005normal}, solid state systems \cite{Reithmaier2004,yoshie2004vacuum,Wallraff2004,chiorescu2004coherent} and an ensemble of trapped ions \cite{herskind2009realization}. 
	However, it has thus far remained elusive for single trapped ions. 
	
	Due to their outstanding properties such as stationary trapping, long coherence times \cite{PhysRevLett.113.220501} and the ability of high-fidelity quantum control \cite{PhysRevLett.117.060504}, trapped ions are a leading system for optical atomic clocks \cite{chou2010frequency,huntemann2016single}, quantum metrology\cite{kotler2011single,baumgart2016ultrasensitive}, and quantum simulation \cite{zhang2017observation} and computation \cite{Debnath2016,monz2016realization}.
	The setting of cavity QED brings about exciting possibilities to connect individual quantum devices by providing efficient reversible quantum interfaces with optical photons \cite{kimble2008quantum}. This enables the distributed architecture using photonic quantum networks for large-scale quantum information processing using trapped ions \cite{PhysRevA.89.022317}. Here we demonstrate a key for this enabling technology by coupling a single ion to an optical cavity in the strong coupling regime for the first time.
	
	In the past, conventional Fabry-Perot cavities with macroscopic mirrors were successfully combined with ion traps to realize a stable single photon source \cite{Keller:04}, tunable entanglement \cite{stute2012tunable} and state transfer between single ions and photons \cite{stute2013quantum}.
	In these experiments, however, the coherent ion-cavity coupling was in the weak coupling regime due to the macroscopic sizes of the optical cavities.
	Since the emitter-cavity coupling scales as $\propto 1/\sqrt{V_m}$ where $V_m$ is the mode volume of the cavity, it is essential to reduce $V_m$ to achieve strong coupling. 
	The main challenge in ion-cavity systems is to achieve small enough mode volume without disturbing the trapping field when incorporating dielectric cavity mirrors near the trapping region.
	In this regard employing laser machined fiber-based Fabry-Perot cavities (FFPCs) has proven to be a viable solution and resulted in several successful implementations recently \cite{Steiner:13,Ballance:16,PhysRevA.96.023824}. Based on the ion trap with an integrated FFPC presented in \cite{PhysRevA.96.023824,kassa2017precise}, in this work we achieve a coherent ion-cavity coupling
	of $g = 2\pi\times (12.3 \pm 0.1)$ MHz greater than both atomic decay rate of the $P_{1/2}$ state of $\gamma = 2\pi\times 11.5$ MHz \cite{Hettrich2015} and cavity decay rate of $\kappa = 2\pi\times (4.1 \pm 0.1)$ MHz. 
	\begin{figure}[tb]
		\begin{center}
			\includegraphics[width=\linewidth]{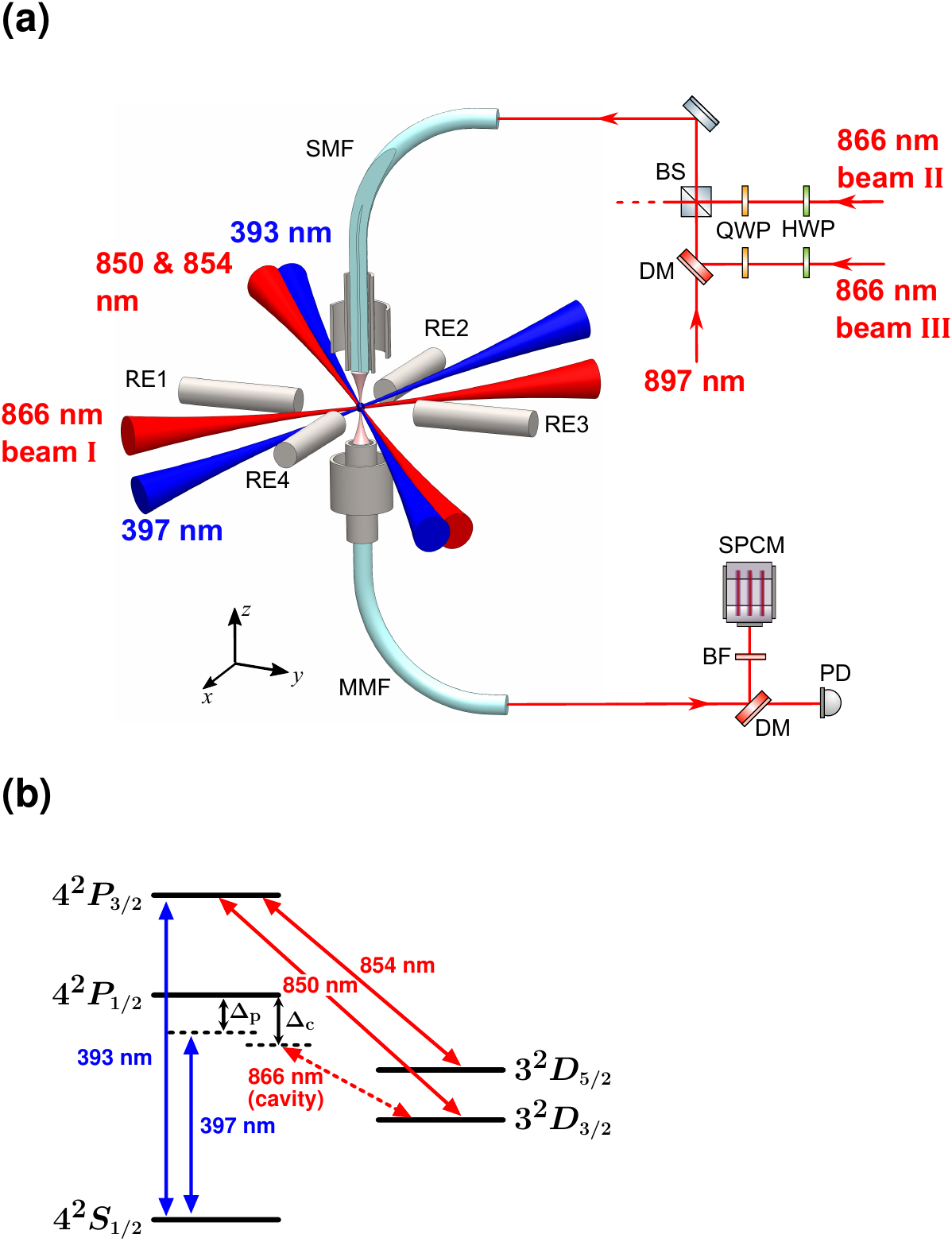}
			\caption{{\bf (a)} A schematic of the experimental setup. A cross-section of the upper trap assembly is shown to reveal the internal structure. BF: bandpass filter, BS: beam splitter, DM: dichroic mirror, HWP: half wave plate, MMF: multimode fiber, PD: photodiode, QWP: quarter wave plate, RE: radial electrode, SMF: single-mode fiber, SPCM: single photon counting module. {\bf (b)} Energy level diagram of \Ca ion with driving lasers and the cavity on the relevant transitions.}
			\label{fig:setup}
		\end{center}
	\end{figure}
	
	Our main experimental apparatus is an endcap-style Paul trap for \Ca ions with an integrated FFPC as shown in \figref~\ref{fig:setup}(a). The main part of the trap is a pair of electrode assemblies, each consisting of two concentric electrodes separated by a ceramic spacer. 
	A radio-frequency (rf) voltage at 19.8\,MHz is applied to the outer electrodes while the inner electrodes are held at rf-ground to create a trapping quadrupole electric field in between the assemblies. 
	Two optical fibers whose end facets act as cavity mirrors are inserted inside the inner electrodes. As a result, a Fabry-Perot cavity is formed in the gap between the assemblies (see \figref\ref{fig:setup}). 
	This design enables effective shielding of the dielectric surfaces of the fibers, and thus achieves robust trapping of a single ion inside an optical cavity with a small mode volume.
	One of the fibers is a single-mode fiber (SMF) to serve as a cavity input whereas the other one is a multimode fiber (MMF) to collect the cavity output. The FFPC mirrors are coated for a wavelength of 866$\,$nm in order to be coupled to the electronic transition between $P_{1/2}$ and $D_{3/2}$ states of $^{40}\mathrm{Ca}^+$. The cavity field decay rate $\kappa$ is measured to be $2\pi\times (4.1\pm 0.1)\,$MHz \cite{Supplementary}.
	With a cavity length of 370$\,\mu$m, this corresponds to a cavity finesse of 50000. 
	
	In addition to the main trapping electrodes, four cylindrical electrodes are radially placed at right angles at a distance of approximately 1 mm from the ion. Dc voltages are applied to two of these radial electrodes (RE1 and RE2 in \figref~\ref{fig:setup}(a)) in orthogonal directions as well as the inner electrodes of the main assemblies to compensate stray electric fields and null the ion's excess micromotion. 
	The other two radial electrodes (RE3 and RE4) are used to displace the rf potential minimum by applying signals synchronous and in-phase to the main drive \cite{kassa2017precise}. In this way the ion is translated radially without incurring excess micromotion and its overlap with the cavity field is optimized. Using a trapped ion as a probe for the cavity field distribution, as demonstrated in \cite{kassa2017precise}, we determine that the center of the $\mathrm{TEM}_{\mathrm{00}}$ cavity mode is located at ($3.4\pm0.1$, $6.4\pm0.3$)$\,\mu$m in the $x$ and $y$ directions respectively from the ion's original position when no additional rf signals are applied to the radial electrodes.
	
	The ion is irradiated by a number of lasers as shown in \figref\ref{fig:setup}(a) with each of them addressing a specific transition of \Ca (see \figref\ref{fig:setup}(b)).
	The ion is Doppler cooled on the $S_{1/2}-P_{3/2}$ transition with a laser at 393\,nm to circumvent inefficient cooling on the the $S_{1/2}-P_{1/2}$ transition caused by the back action of the strong Purcell effect when the cavity is near resonant on the $P_{1/2}-D_{3/2}$ transition \cite{PhysRevA.96.023824}.
	Lasers at 850\,nm and 854\,nm repump the ion from the meta-stable $D$ states into the $S_{1/2}$ state for continuous cooling.
	Three laser beams at 866\,nm, denoted as beam I, II and III, with individual polarization controls are used for optical pumping and probing of the ion. Two of them (beam II and III) are injected into the input SM fiber to drive the FFPC. A laser beam at 897\,nm is also sent into the FFPC through the SM fiber. Its transmission is filtered out from the main cavity signal at the output of the MM fiber and used to stabilize the length of the FFPC.  
	
	\begin{figure}[tb]
		\begin{center}
			\includegraphics[width=\linewidth]{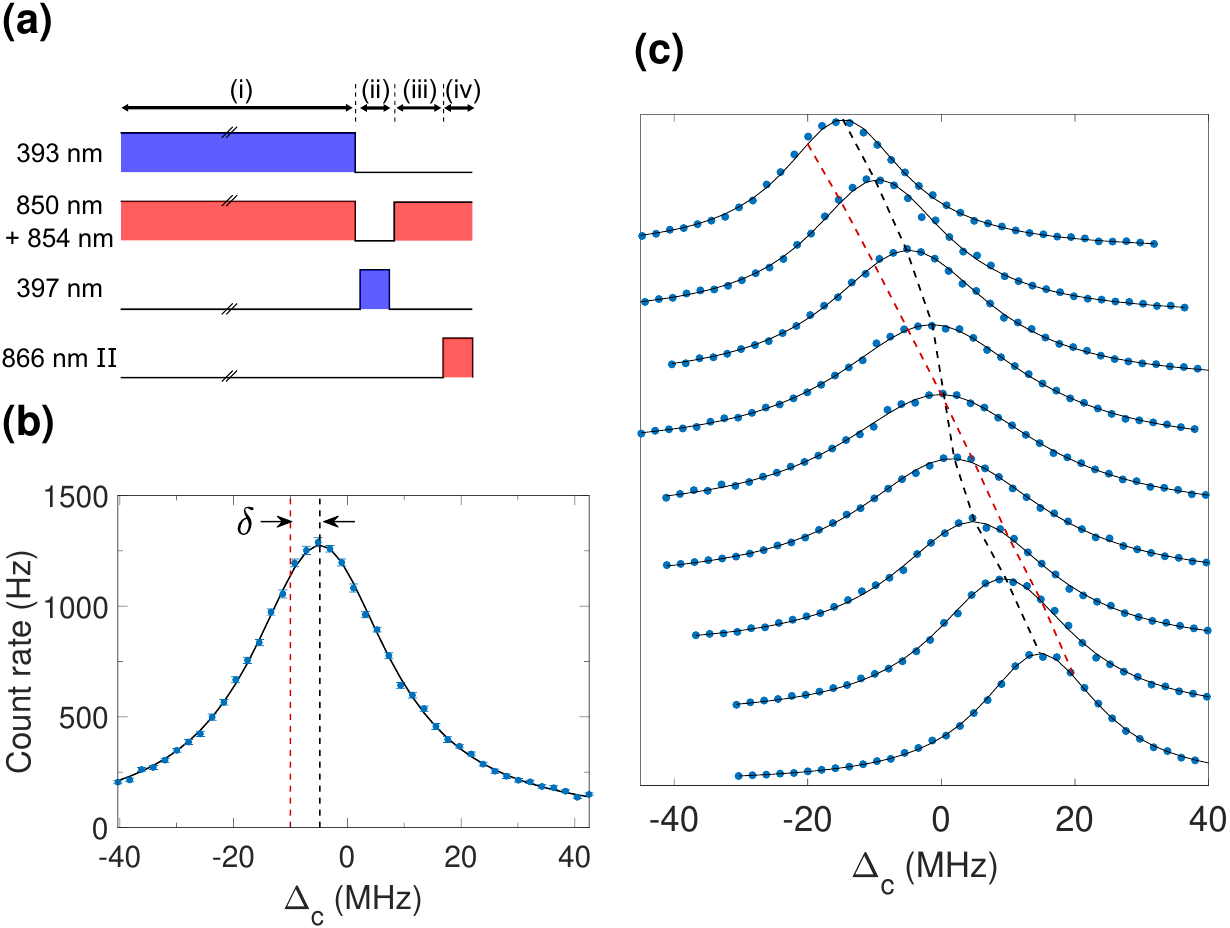}
			\caption{{\bf(a)} The pulse sequences for the single-photon generation via the vacuum-stimulated Raman transition: (i) Doppler cooling for 6\,$\mu$s. (ii) A 300\,ns-long pulse of the 397\,nm laser is applied producing a single photon in the cavity. (iii) Recycling the ion's population back to the $S_{1/2}$ state for 500\,ns. (iv) A pulse of the 866\,nm laser is injected to the cavity through the input fiber. As the 866\,nm laser is frequency-locked to the exact resonance to the $P_{1/2}-D_{3/2}$ transition, a peak from the 866\,nm laser pulse  provides an absolute frequency reference for $\Delta_c$. The SPCM counts during (ii) and (iv) are independently measured. {\bf(b)} Single-photon emission spectrum as a function of $\Delta_c$ with a fixed $\Delta_p$ at -10\,MHz. The solid line is a fit by the Lorentzian function. The vertical dashed lines indicate the center frequency of the peak (black) and the frequency expected from the condition $\Delta_p = \Delta_c$ (red). The same applies to the dashed lines in (c). {\bf(c)} A stacked plot of single-photon emission spectra with different $\Delta_p$. From the top to the bottom traces, $\Delta_p$ varies from -20 to +20\,MHz with an interval of 5 MHz.}
			\label{fig:cavityscan}
		\end{center}
	\end{figure}
	
	\begin{figure}[h]
		\begin{center}
			\includegraphics[width=0.8\linewidth]{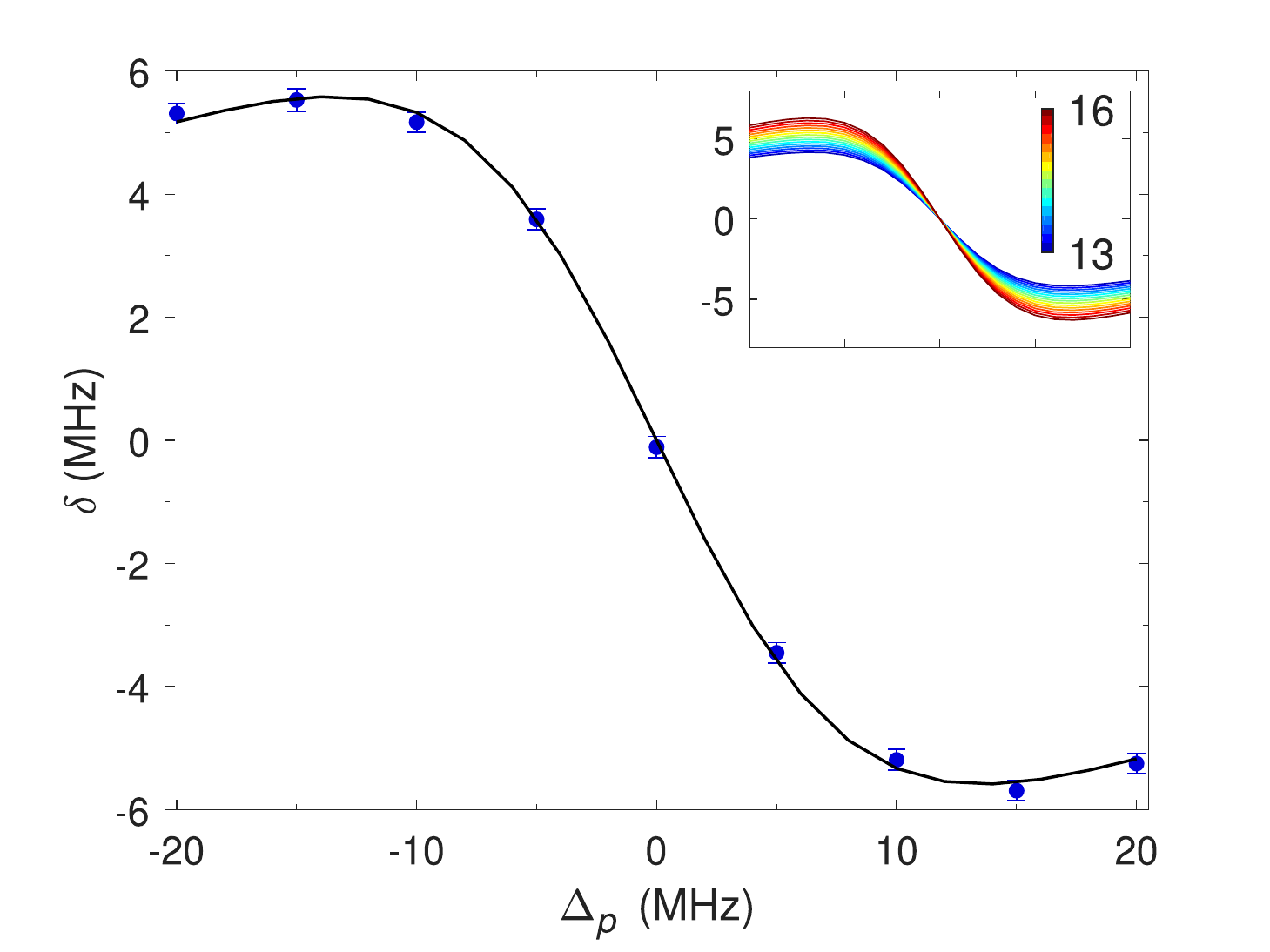}
			\caption{The shift of the Raman resonance $\delta$ as a function of $\Delta_p$ from the data set in \figref\ref{fig:cavityscan}(c). The error bars are the statistical mean standard errors. The solid line is a fit by the numerical simulation. The inset figure shows superimposed traces of $\delta$ from numerical simulations with different $g_0/(2\pi)$. The variation of $g_0/(2\pi)$ from $13$ to $16$\,MHz is represented by the gradation of the line colors. The horizontal range of the inset is identical to that of the main figure.}
			\label{fig:delta_vs_Deltap}
		\end{center}
	\end{figure}
	
	Previously a coherent ion-cavity coupling $g_0$ of $2\pi\times (5.3\pm 0.1)\,$MHz was measured in the same trap without optimizing the radial position of the ion \cite{PhysRevA.96.023824}. Having moved the ion to the radial center of the FFPC, we now characterize $g_0$ with the optimized overlap at the anti-node of the cavity field.
	The ion-cavity coupling can be quantified by analyzing the single-photon emission spectra of the ion-cavity system.
	\figref\ref{fig:cavityscan}(a) shows the pulse sequences of the lasers for this measurement. 
	After a period of Doppler cooling in (i), a short pulse of the 397 nm laser with a detuning $\Delta_p$ is applied in (ii). 
	In combination with the cavity locked close to the $P_{1/2}-D_{3/2}$ transition with a detuning $\Delta_c$, this results in the production of a single photon in the cavity via a vacuum-stimulated Raman transition from the $S_{1/2}$ to $D_{3/2}$ state \cite{Keller:04}. The single photon emission is followed by the optical pumping of the ion into the $S_{1/2}$-state in (iii) and the probing of the cavity frequency without ion-cavity interaction in (iv).

	Normally the Raman resonance condition for the single photon emission dictates $\Delta_p = \Delta_c$.
	However, due to the quantum mechanical coupling between the ion and cavity, this resonance condition is shifted \cite{albert2012collective}. \figref\ref{fig:cavityscan}(b) shows a spectrum of detected single photons from the FFPC as a function of $\Delta_c$ while $\Delta_p$ is fixed (-10\,MHz in this particular case).
	It can be clearly seen that the peak frequency of the spectrum (black dashed line) is shifted by an amount $\delta$ from the expected $\Delta_p = \Delta_c$ condition (red dashed line).
	We repeat this Raman spectroscopy for different $\Delta_p$ as shown in \figref\ref{fig:cavityscan}(c) in order to measure the dependence of $\delta$ on $\Delta_p$. The frequency shift $\delta$ exhibits a dispersion-like profile whose amplitude and gradient depend on the magnitude of $g_0$.
	Because $\delta$ also depends on the Rabi frequency $\Omega_{\mathrm{397}}$ of the 397\,nm laser through its own ac Stark shift, we independently measure $\Omega_{\mathrm{397}}$ to be $2\pi\times(11.9 \pm 0.4)$\,MHz by the electron shelving method as employed in \cite{PhysRevA.96.023824}.
	Given $\Omega_{\mathrm{397}}$ and other known experimental parameters such as the beam detunings, beam polarizations and the magnetic field, the single-photon emission spectrum and hence $\delta$ can be precisely simulated based on the 8-level model of $^{40}\mathrm{Ca}^+$ with $g_0$ as the only free parameter (see the inset of \figref~\ref{fig:delta_vs_Deltap}).
	Utilizing the dependence of $\delta$ on $g_0$ and fitting this numerical model to the experimental data as shown in \figref~\ref{fig:delta_vs_Deltap}, we obtain the coherent ion-cavity coupling $g_0 = 2\pi\times (15.1\pm0.1)$\,MHz. The uncertainty of $g_0$ includes the error of the fit as well as the contributions from the uncertainties of the model parameters in the numerical simulation \cite{Supplementary}.
	
	\begin{figure}[t]
		\begin{center}
			\includegraphics[width=\linewidth]{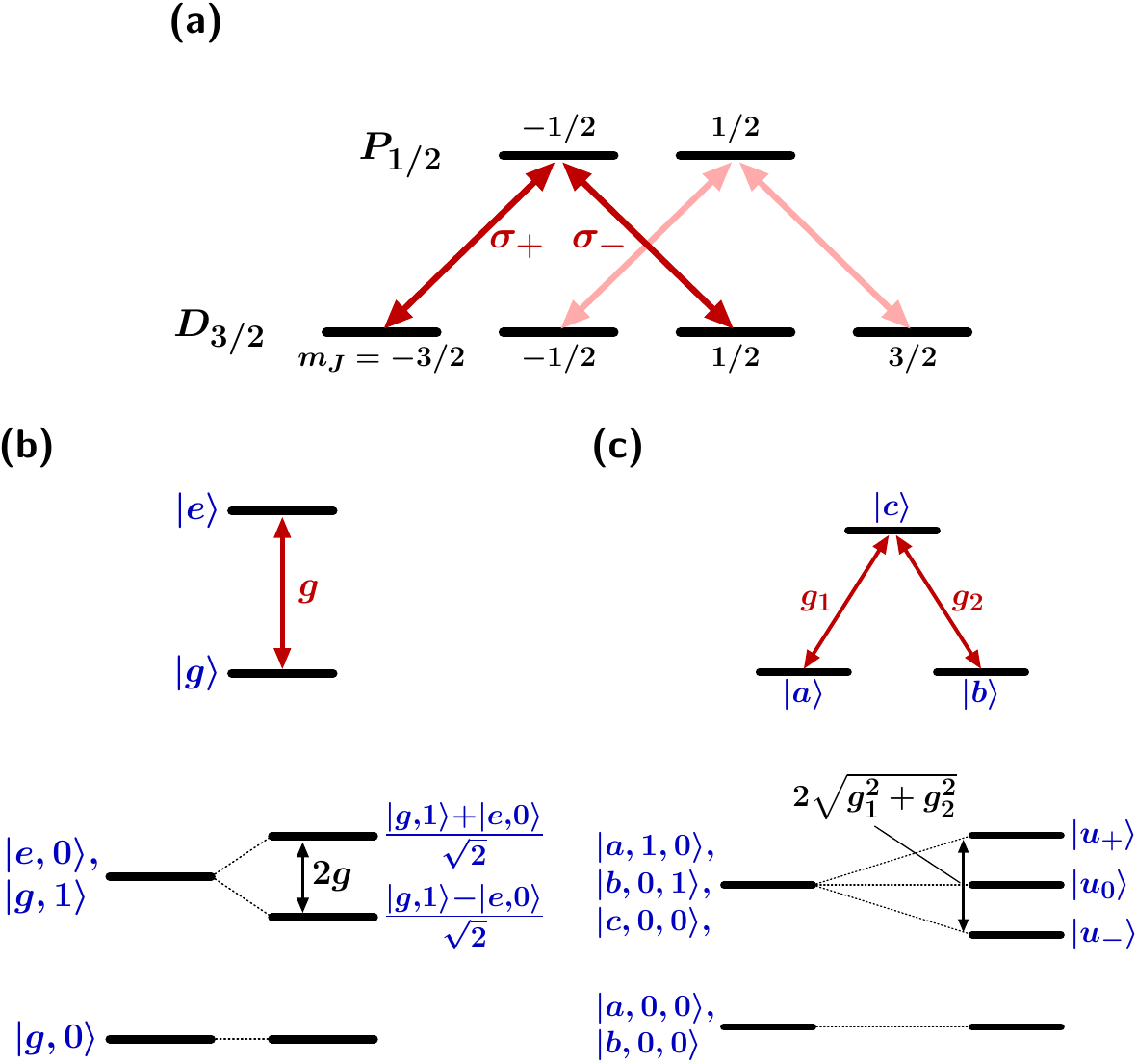}
			\caption{{\bf (a)} The Zeeman sublevels in the $P_{1/2}$ and $D_{3/2}$ state manifolds. The two polarization modes in the FFPC couple to the $\sigma_+$ and $\sigma_-$ transitions respectively. {\bf (b)} Top: A two-level atom coupled to a resonant optical single mode with a coupling strength $g$. Bottom: The level diagram of the total energy  of the system with/without the atom-cavity coupling (right/left). {\bf (c)} Top: A three-level atom coupled to two optical modes simultaneously with coupling strength $g_1$ and $g_2$ respectively. Bottom: The level diagram of the three-level bimodal system with/without the atom-cavity coupling (right/left). The energy level of the first excited states split into three levels with corresponding dressed states.}
			\label{fig:sublevel_schemes}
		\end{center}
	\end{figure}
	
	\begin{figure}[!t]
		\begin{center}
			\includegraphics[width=0.9\linewidth]{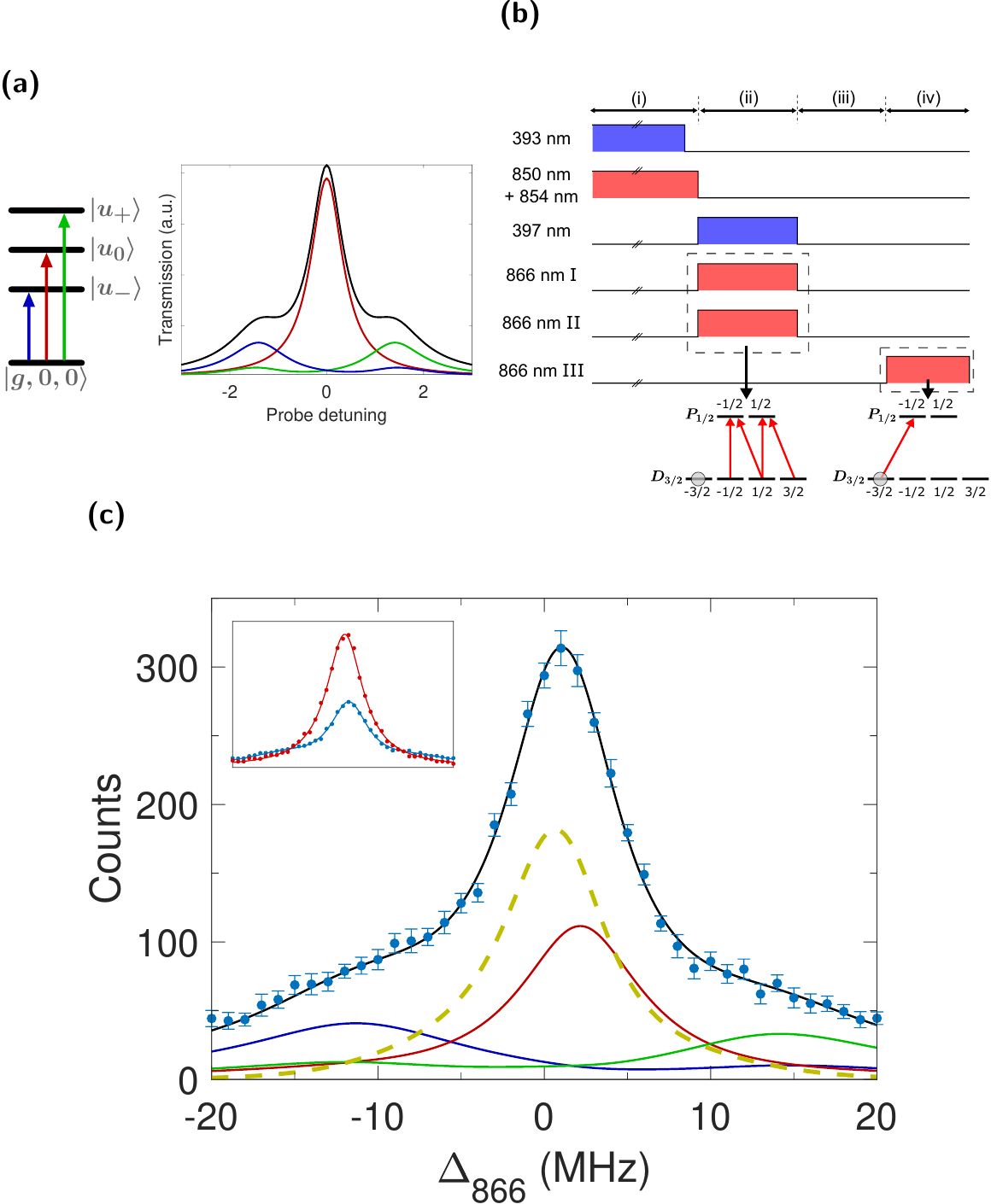}
			\caption{{\bf (a)} A model calculation for the ideal three-level system. The solid line shows the expected spectrum of transmitted photons as a function of the probe detuning. Here $g_1=g_2=g$ and the probe frequency is normalized by $g$. The underlying contributions of the individual dressed states are shown in the same colors as the corresponding  excitations in the level diagram on the left. {\bf (b)} Laser pulse sequences. (i) 5\,$\mu$s-long Doppler cooling. The duration of the repumping beams is longer than that of the 393\,nm beam in order to prepare the ion in the $S_{1/2}$ state at the end of the cooling. (ii) Optical pumping for 3\,$\mu$s. (iii) An interval is placed in order to wait for the intensities of the optical pumping lasers to sufficiently diminish. (iv) Probing with the 866\,nm beam III. {\bf(c)} The counts of the transmitted photons of 866\,nm beam III as a function of its detuning. The background counts from stray light are subtracted. The solid black line is the result of numerical calculation (see the main text). The underlying contributions are also shown with the same color scheme as in (a). In addition the contribution from non-dressed states is shown in the yellow dashed line. The inset also shows a spectrum taken without the ion (red) superposed with the spectrum with the ion (blue).}
			\label{fig:trans-spec}
		\end{center}
	\end{figure}
	
	A small magnetic field (= 0.9 gauss) is applied to align the quantization axis to the cavity axis such that the cavity supports two distinct polarizations $\sigma_+$ and $\sigma_-$. As shown in \figref~\ref{fig:sublevel_schemes}(a), the ion is simultaneously coupled to these two polarization modes on the transitions connecting the Zeeman sublevels in the $P_{1/2}$ and $D_{3/2}$ state manifolds. As far as the coherent ion-cavity interaction is concerned, this configuration effectively realizes a closed three-level lambda system interconnected via a bimodal cavity. 
In the standard cavity QED model where a two-level atom is coupled to a single optical mode, there are two dressed states $(\ket{g,1}+\ket{e,0})/\sqrt{2}$ and $(\ket{g,1}-\ket{e,0})/\sqrt{2}$ with an energy gap $2g$ ($\hbar = 1$) in the subspace corresponding to the first excitation from the ground state ($=\ket{g,0}$) (see \figref\ref{fig:sublevel_schemes}(b)). Here $g$ and $e$ denote the ground and excited states of the atom, and $0$ and $1$ denote the intracavity photon number. As a result, a coherent oscillation between $\ket{g,1}$ and $\ket{e, 0}$ occurs at the vacuum Rabi frequency of $2g$.
	Similarly, for the bimodal system with three atomic levels, the subspace for the first excitation includes three originally degenerate states $\ket{a,1,0}$, $\ket{b,0,1}$ and $\ket{c,0,0}$. Here the notation indicates a product of the atomic state and the photon number states of the two cavity modes (see \figref\ref{fig:sublevel_schemes}(c) for the labeling of the atomic levels). Due to the atom-cavity coupling, the system now has the following three dressed states:
	\begin{align}
	\ket{u_+} &= \frac{g_1\ket{a,1,0}+g_2\ket{b,0, 1}+\lambda\ket{c,0,0}}{\sqrt{2}\lambda}, \\
	\ket{u_0} &= \frac{g_2\ket{a,1,0}-g_1\ket{b,0,1}}{\lambda},\\
	\ket{u_-} &= \frac{g_1\ket{a,1,0}+g_2\ket{b,0, 1}-\lambda\ket{c,0,0}}{\sqrt{2}\lambda},
	\end{align}
	where $\lambda = \sqrt{g_1^2+g_2^2}$.
	%
	Note that $\ket{u_0}$ is a dark state which is decoupled from the excited atomic upper state $\ket{c}$. The emergence of this state is very similar to the effect of electromagnetically induced transparency (EIT) \cite{fleischhauer2005electromagnetically}. The difference here is that the quantized cavity fields, instead of classical lasers, interconnect the three atomic levels. While in \cite{tanji2011vacuum}, one of the EIT transitions was replaced by a quantized cavity field in our case both of them are taken up by the cavity fields.
	On the other hand, a \textit{bright state} can also be constructed as $\ket{v} = (g_1\ket{a,1,0}+g_2\ket{b,0,1})/\lambda$ in which the excitation amplitudes to $\ket{c,0,0}$ from the constituent states interfere constructively. Using $\ket{v}$, 
	\begin{align}
 \ket{u_\pm} = \frac{\ket{v}\pm\ket{c,0,0}}{\sqrt{2}},
	\end{align}
and the energy gap between $\ket{u_+}$ and $\ket{u_-}$ is 2$\lambda$ 
as seen in \figref\ref{fig:sublevel_schemes}(c).
Consequently, in the same way as between $\ket{g,1}$ and $\ket{e,0}$ in the two-level case, the coherent oscillation occurs between $\ket{v}$ and $\ket{c,0,0}$ at the frequency of 2$\lambda$. This oscillation corresponds to the characteristic emission and absorption of a single photon in this system where the single photon is emitted into and absorbed from the two optical modes simultaneously in a superposition. Hence the vacuum Rabi frequency --the frequency at which a single excitation is exchanged between the atomic and optical degrees of freedom-- is given by $2g = 2\lambda$.
	
	Applying this picture to the actual energy levels of $^{40}\mathrm{Ca}^+$ in \figref\ref{fig:sublevel_schemes}(a), $g_1$ and $g_2$ are derived from $g_0$ by multiplication with the Clebsh-Gordan coefficients for the $\sigma_+$  and $\sigma_-$ transitions, which are $1/\sqrt{2}$ and $1/\sqrt{6}$ respectively.
	With $g_0 = 2\pi\times (15.1\pm0.1)$\,MHz, $g = 2\pi\times (12.3\pm0.1)$\,MHz is obtained. Therefore the coupling of the single ion to the optical cavity $g$ exceeds both the atomic decay rate of the $P_{1/2}$ level $\gamma$ (=$2\pi\times11.5$\,MHz) \cite{Hettrich2015} and the cavity decay rate $\kappa$ (=$2\pi\times(4.1\pm0.1)$\,MHz), placing our system in the strong coupling regime ($g > \gamma, \kappa$).
	
	The characteristic vacuum Rabi splitting in the three-level bimodal system as shown in \figref~\ref{fig:sublevel_schemes}(c) can be probed by weakly driving the cavity and detecting the transmission, in the same way as in two-level atoms \cite{boca2004observation,maunz2005normal}. 
	\figref~\ref{fig:trans-spec}(a) shows the expected spectrum of the transmitted photons when the ideal three-level bimodal system is probed with a near-resonant coherent light.
	There are three underlying resonant peaks in the resulting spectrum, each corresponding to the three distinct excitations from the ground state. The one corresponding to the EIT state is most prominent in the middle while the excitations to the upper and lower dressed states create the shoulders beside it.
	
	\figref\ref{fig:trans-spec}(b) shows the laser pulse sequences used to probe this structure in the experiment. After a period of Doppler cooling, the ion is optically pumped to an extremal Zeeman state of the $D_{3/2}$ manifold ($m_J = -3/2$). In order to do so, the free space (beam I) and one of the cavity-coupled 866\,nm lasers (beam II) are applied with $\pi$ and $\sigma_-$ polarizations respectively. Subsequently a pulse of 866\,nm beam III in the $\sigma_+$ polarization is injected and its transmission through the FFPC is measured. Prior to the measurement, the ion is moved to the antinode in the cavity field and the FFPC is locked to the atomic resonance ($\Delta_c = 0$). The intensity of 866\,nm beam III in the cavity is estimated in terms of the displacement amplitude to the intra-cavity field \cite{Supplementary}. \figref\ref{fig:trans-spec}(c) shows the resulting spectrum of the transmitted photons as the detuning of the 866\,nm beam III from the atomic resonance ($\equiv\Delta_{866}$) is scanned. The spectrum is significantly modified by the ion-cavity coupling (see the inset of \figref\ref{fig:trans-spec}c).
	The data shows good agreement with the numerical simulation shown as the black solid line in \figref\ref{fig:trans-spec}(c). 
	Only the vertical scaling and a small horizontal offset ($\sim 0.47$ MHz) are adjusted to fit the simulated curve to the measured data.
	The horizontal offset is likely to have resulted from an error in the calibration of the frequency of the 866\,nm laser.
	The asymmetry of the spectrum is due to the applied small magnetic field.
	In the figure also shown are the simulated contributions of the excitations to the individual dressed states and contribution from other states. It is clearly seen that the peaks corresponding to $\ket{u_{\pm}}$ are well separated, exhibiting the vacuum Rabi splitting of the system. 
	On the other hand there is a finite probability that the probing laser excites the ion and incoherently distributes its population via spontaneous decays from the $P_{1/2}$ state. This results in transmission of subsequent photons without interacting with the ion and creates the central peak in a dashed yellow line in the figure. Note that this probability increases as $g$ increases and hence progressively fewer photons are required to probe the system, whereas in practice a certain number of photons are still required at the detector to ensure a decent signal-to-noise ratio.
	
	In conclusion, we have developed a miniaturized ion trap which integrates an FFPC and couples a single ion to its cavity modes with an optimized overlap. We have achieved the strong coupling regime for the first time with a single ion, where the vacuum Rabi frequency of the system exceeds both atomic and cavity decoherence rates. Moreover the characteristic energy structure of the dressed-states inherent to our coupled ion-cavity system has been successfully probed by a spectroscopic means.
Strong coupling between a single ion and an optical cavity facilitates novel opportunities to combine the unparalleled capabilities of trapped ions with quantum photonics. In particular, it enables applications such as highly efficient single photon sources and  high fidelity ion-photon quantum interfaces, key components in quantum networks and quantum computing.

\begin{acknowledgments}
We gratefully acknowledge support from EPSRC through the UK Quantum Technology Hub: NQIT - Networked Quantum Information Technologies (EP/M013243/1 and
EP/J003670/1). 
\end{acknowledgments}

	\bibliographystyle{unsrt}
	\bibliographystyle{apsrev4-1}
	\bibliography{StrongCoupling}

\newpage
\clearpage

\onecolumngrid
\appendix
\begin{center}
{\Large\bf{Supplementary Materials}}
\end{center}

\renewcommand{\thefigure}{S\arabic{figure}}
\renewcommand{\thetable}{S\arabic{table}}
\renewcommand{\theequation}{S\arabic{equation}}
\setcounter{figure}{0}
\setcounter{equation}{0}

\section*{Cavity linewidth}

\begin{figure}[htb]
 \centering
  \includegraphics[width=0.5\linewidth]{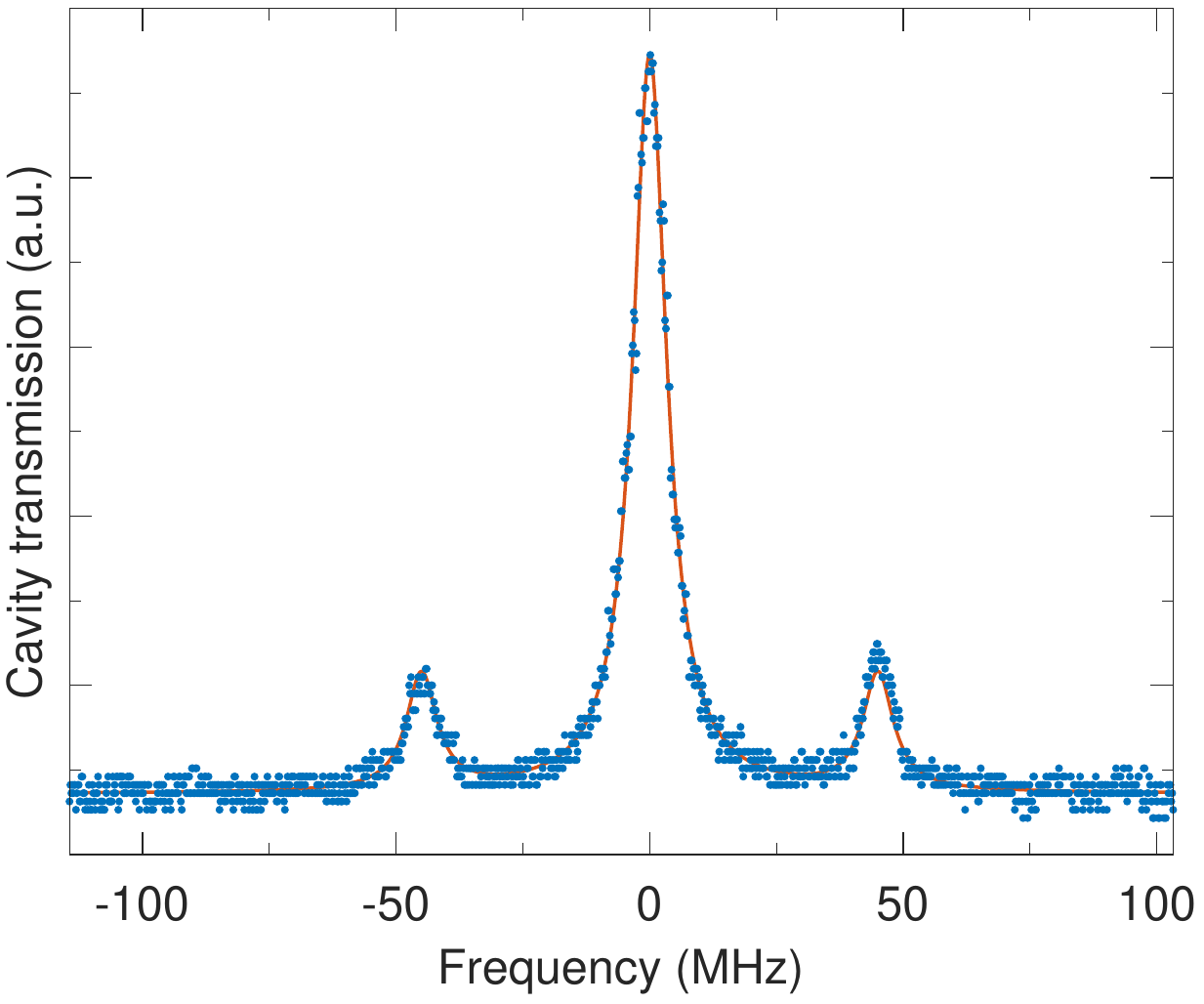}
  \caption{Transmission signal as the FFPC is scanned over resonance. The x-axis is calibrated in frequency by the sideband peaks at $\pm45$\,MHz. The solid line is a fit with Lorentz functions.}
  \label{fig:finesse}
\end{figure}

\figref\ref{fig:finesse} shows a transmission signal when a laser at 866\,nm is injected to the FFPC and the cavity length is scanned over a resonance. The laser is modulated at 45\,MHz such that the sidebands on the transmission signal serve as a frequency reference. From the fit, the half-width-half-maximum of the central peak ($=\kappa/(2\pi)$) is deduced. Repeating this measurement 20 times, we obtained $\kappa = 2\pi\times (4.1\pm 0.1)\,$MHz corresponding to a cavity finesse of $\sim$50000. 

\section*{Theoretical model}

The system Hamiltonian in the Raman spectroscopy in an interaction picture is described as follows:
\begin{align}
 H &= H_0 + H_B + H_\mr{pump} + H_\mr{ion-cav}, \label{eq:hamiltonian}
\end{align}
with the atomic detuning term
\begin{align}
 H_0 &= \Delta_p \sum_{m_S} \proj{S,m_S}{S,m_S} + \Delta_c \sum_{m_D} \proj{D,m_D}{D,m_D}, \label{eq:H0}
\end{align}
the Zeeman term
\begin{align}
 H_B &= B\sum_{L = S,P,D}\sum_{m_j} g_L \mu_B m_j \proj{L,m_j}{L,m_j}, \label{eq:HB}
\end{align}
the pump term describing the interaction of the ion with the pump laser at the $S_{1/2}\rightarrow P_{1/2}$ transition,
\begin{align}
 H_\mr{pump} &= \frac{\Omega_\mr{397}}{2}f(t)\sum_q\sum_{m_S, m_P}\left[\eps_qC(j_Sm_S,1q;j_Pm_P)\proj{P, m_P}{S, m_S} + \mr{H.c.}\right],\label{eq:Hpump}
\end{align}
and the ion-cavity interaction,
\begin{align}
 H_\mr{ion-cav} &= g_0\sum_{m_P, m_D}\left[C(j_Dm_D,11;j_Pm_P)a_+\proj{P, m_P}{D, m_D} + C(j_Dm_D,1 -1;j_Pm_P)a_-\proj{P, m_P}{D, m_D} + \mr{H.c.}\right].\label{eq:Hcavity}
\end{align}
Here $\hbar = 1$. The atomic part of the state is described by the Zeeman states $\{\ket{S, m_S}, \ket{P, m_P}, \ket{D, m_D}\}$. $B$ is the amplitude of the applied magnetic field, $g_L (L = S,P,D)$ is the Land\'{e} g-factor, $\mu_B$  is the Bohr magneton, $\epsilon_q$ is the polarization component of the pump beam in the spherical basis, and $C(j_Um_U,1q;j_Vm_V)$ is the Clebsch-Gordan coefficient given by a Wigner 3-j symbol:
\begin{align}
 C(j_Um_U,1q;j_Vm_V) &= (-1)^{j_U-1+m_V}\sqrt{2j_V+1}\begin{pmatrix}
  j_U & 1 & j_V\\
  m_U & q  & -m_V\\
\end{pmatrix}.
\end{align}
$a_+$ and $a_-$ are the annihilation operators for the $\sigma_+$ and $\sigma_-$ polarization modes of the cavity. $f(t)$ represents the pulse shape of the pump beam.
Numerical simulation are carried out by solving master equations given by,
\begin{align}
 \frac{\pt\rho}{\pt t} = -i[H, \rho]+\kappa\mc{L}(\rho, a_+)+&\kappa\mc{L}(\rho, a_-)+ \gamma_S\sum_{m_S, m_P, q}\mc{L}(\rho, C(j_Sm_S,1q;j_Pm_P)\proj{S, m_S}{P, m_P}) \nonumber\\
 &+ \gamma_D\sum_{m_D, m_P, q}\mc{L}(\rho, C(j_Dm_D,1q;j_Pm_P)\proj{D, m_D}{P, m_P}),\label{eq:master-eq}
\end{align}
where $\mc{L}(\rho, \mc{O}) \equiv 2\mc{O}\rho\mc{O}^\dag-\mc{O}^\dag\mc{O}\rho-\rho\mc{O}^\dag\mc{O}$, and $\gamma_S$ and $\gamma_D$ are the atomic decay rates from the $P$ state to the $S$ and $D$ states respectively. All the numerical simulations are carried out using the Quantum Optics toolbox for MATLAB \cite{Tan:99}.

\section*{Uncertainty of $g_0$}

\begin{figure}[htbp]
 \centering
  \includegraphics[width=0.5\linewidth]{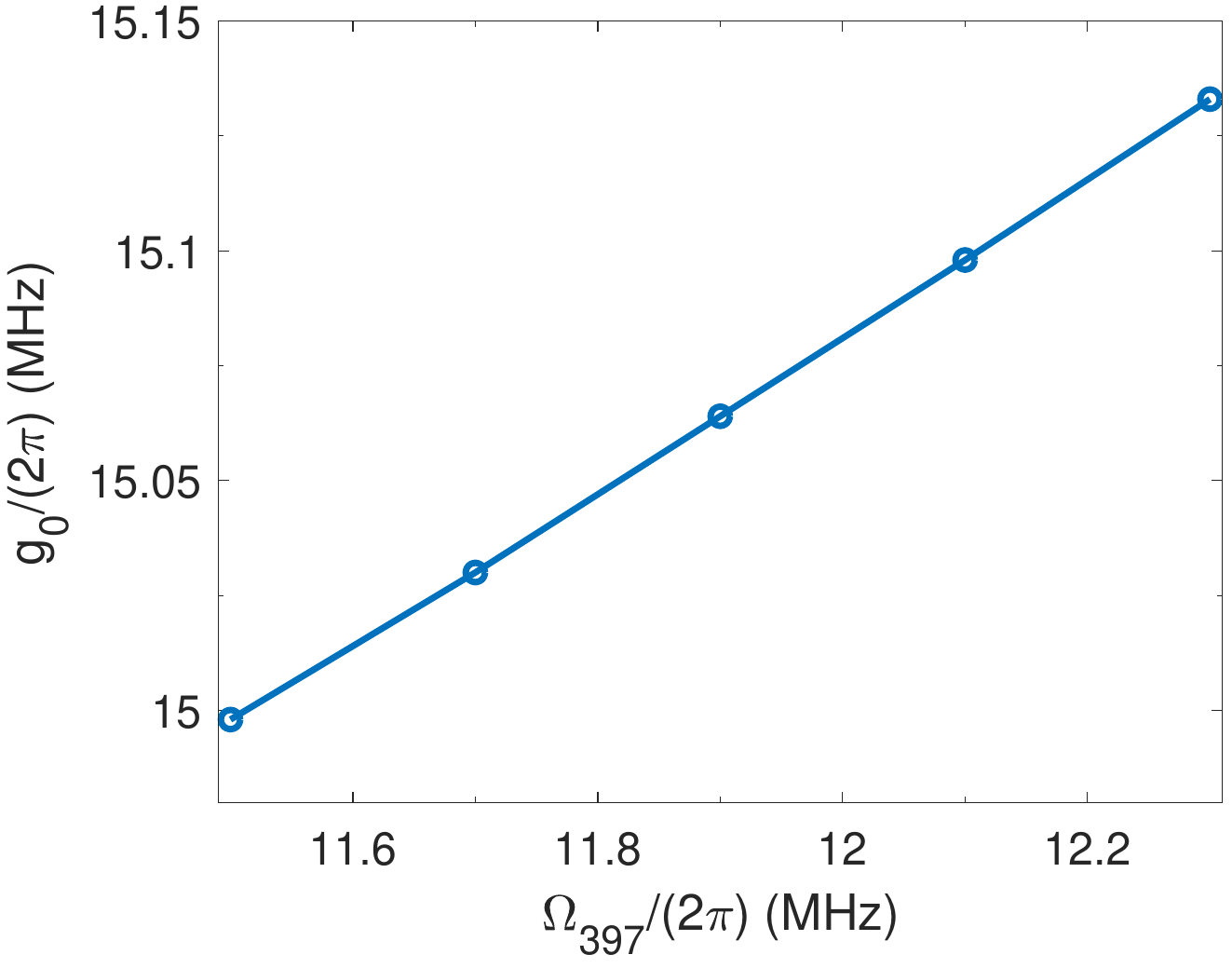}
  \caption{Numerically obtained dependence of $g_0$ on $\Omega_\mr{397}$ around $\Omega_\mr{397} = 2\pi \times11.9$\,MHz. From the gradient, we obtain the sensitivity of $g_0$ on the model parameter $\Omega_\mr{397}$.}
  \label{fig:Omega397vsg0}
\end{figure}

The fit shown in \figref3 is obtained by fitting the solution of Eq.(\ref{eq:master-eq}) to the data with $g_0$ as a fitting parameter. However uncertainties in other physical parameters such as $\Omega_\mr{397}$ affect the resultant uncertainty of $g_0$. Therefore we study the effects of the model parameters by varying their values and refitting the model to obtain the deviated $g_0$ and find out a change from the original value of $g_0$. We validate that the dependence of $g_0$ on the parameters is linear in the region of our interest (e.g. see \figref\ref{fig:Omega397vsg0}). Through this variational analysis, we propagate the error of each parameter to that of $g_0$ as listed in TABLE~\ref{tab:param-error}. Combining all the error budgets, the uncertainty of $g_0$ from the uncertainties of the model parameters is determined to be 0.07\,MHz. As a result, together with the fitting error of 0.07\,MHz in \figref3, we get $g_0 = 2\pi\times(15.1\pm 0.1)$\,MHz.   

\begin{table}[t]
     \begin{tabular}{|c|c|c|c|}
      \hline
      Parameter & Error of param. & Gradient & Error budget for $g_0$\\
      \hline
      $\Omega_{\mr{397}}$ & $(2\pi)$0.4 MHz & 0.17 & $(2\pi)$ 0.07 MHz \\
      \hline
      $B$ & $\sim$0.1 Gauss & 0.18 MHz/Gauss & $\sim$ $(2\pi)$ 0.02 MHz  \\
      \hline
      $\kappa$ & $(2\pi)$0.1 MHz & 0.07  & $(2\pi)$ 0.01 MHz  \\
      \hline
     \end{tabular}
 \caption{Estimated errors of the model parameters in the numerical simulation and their contributions to the error of $g_0$ obtained through the variational analysis.}
 \label{tab:param-error}
\end{table}

\section*{Expected value of  $g_0$ at the Doppler cooling limit}

From the geometry of the cavity \cite{PhysRevA.96.023824}, the ideal value of $g_0^{(\mr{ideal})}$ is calculated to be $(2\pi)\times 17.3$\,MHz when the ion's positional spread is taken to be zero. In reality the ion has a finite positional spread according to the temperature, the secular frequency of the trap and the beam angle of the 393\,nm laser \cite{itano1982laser}. At the Doppler cooling temperature, the positional spread along the direction of the cavity axis is calculated to be $\Delta z = 94$\,nm. Note that the spread along other two directions are neglected since the spatial variations of the cavity field in these directions are negligible compared to the axial direction. From this, the expected ion-cavity coupling is deduced to be $g_0 = g_0^{(\mr{ideal})}\sqrt{\frac{1+\e^{-k^2\Delta z^2}}{2}} = 2\pi \times 15.6$\,MHz which compares well with the experimentally obtained value of $2\pi \times 15.1$\,MHz.  

\section*{Estimation of the cavity driving amplitude}

For the transmission spectroscopy of Fig.5, the system Hamiltonian is modified in order to incorporate the cavity driving term:
\begin{align}
 H &= H_0' + H_B + H_\mr{drive} + H_\mr{ion-cav}, \label{eq:hamiltonian_trans}
\end{align}
with
\begin{align}
 H_0' &= \Delta_{866} \sum_{m_D} \proj{D,m_D}{D,m_D}-\Delta_{866}(a_+^\dag a_+ + a_-^\dag a_-), \label{eq:H0_trans}
\end{align}
and
\begin{align}
 H_\mr{drive} &= E(a_+ + a_+^\dag). \label{eq:Hdrive}
\end{align}
Here we assume $\Delta_\mr{cav} = 0$.

\begin{figure}[htbp]
 \centering
  \includegraphics[width=0.5\linewidth]{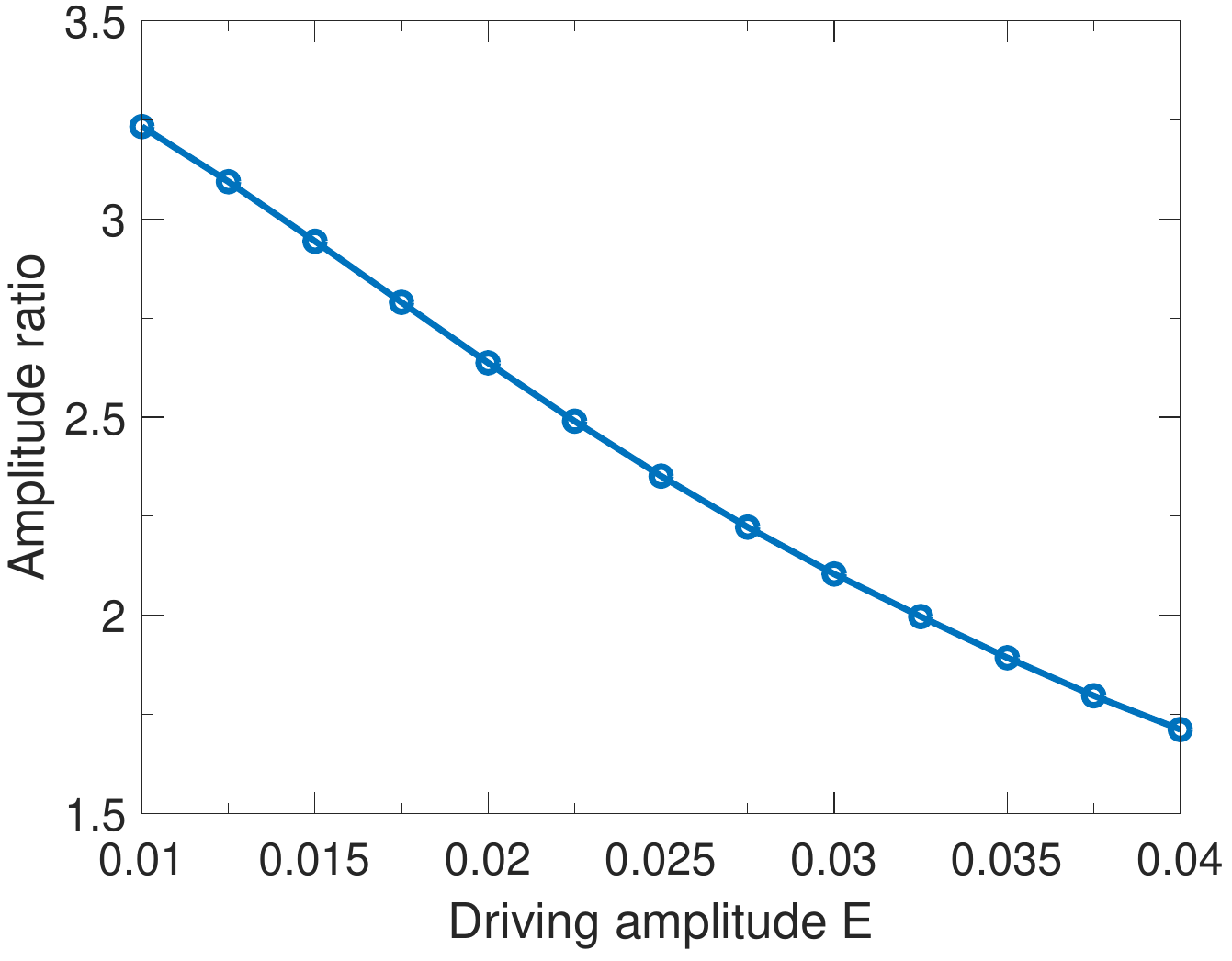}
  \caption{The ratio of the peak heights with and without an ion in the cavity transmission spectroscopy is numerically calculated for different driving field amplitude.}
  \label{fig:amp_ratio_vs_drive}
\end{figure}

In order to estimate the magnitude of $E$ in (\ref{eq:Hdrive}), we use the transmission spectrum taken without an ion (red trace in the inset of Fig.5) as a reference. As $E$ increases, the probability of incoherent scattering increases and the shape of the spectrum taken with the ion asymptotically approaches to the one without the ion. Consequently the ratio between the peak heights of the spectrums with and without the ion depends on the magnitude of $E$ as shown in \figref~\ref{fig:amp_ratio_vs_drive}. Using this dependence, we obtain $E = 0.032$ from the experimental data shown in Fig.~5. 

\end{document}